\documentclass[twocolumn]{aastex61}

\usepackage{CJKutf8}

\begin{document}
\title{X-ray Luminosity and Size Relationship of Supernova Remnants in the LMC}
%
%
\begin{CJK*}{UTF8}{bsmi}
\author{Po-Sheng Ou (歐柏昇)}
\affiliation{Institute of Astronomy and Astrophysics, Academia Sinica, P.O. Box 23-141, Taipei 10617, Taiwan, R.O.C.\ 
\\ psou@asiaa.sinica.edu.tw}
\affiliation{Department of Physics, National Taiwan University, Taipei 10617, Taiwan, R.O.C.}

\author{You-Hua Chu (朱有花)}
\affiliation{Institute of Astronomy and Astrophysics, Academia Sinica, P.O. Box 23-141, Taipei 10617, Taiwan, R.O.C.\ 
\\ psou@asiaa.sinica.edu.tw}
\affiliation{Department of Physics, National Taiwan University, Taipei 10617, Taiwan, R.O.C.}
\affiliation{Department of Astronomy, University of Illinois at Urbana-Champaign, 1002 West Green Street, \\
Urbana,IL 61801, U.S.A.}

\author{Pierre Maggi}
\affiliation{D\'epartement d'Astrophysique, IRFU, CEA, Universit\'e Paris-Saclay, F-91191 Gif-sur-Yvette, France.}

\author{Chuan-Jui Li (李傳睿)}
\affiliation{Institute of Astronomy and Astrophysics, Academia Sinica, P.O. Box 23-141, Taipei 10617, Taiwan, R.O.C.\ 
\\ psou@asiaa.sinica.edu.tw}
\affiliation{Department of Physics, National Taiwan University, Taipei 10617, Taiwan, R.O.C.}

\author{ Un Pang Chang (曾遠鵬)}
\affiliation{Department of Physics, National Taiwan University, Taipei 10617, Taiwan, R.O.C.}

\author{Robert A. Gruendl}
\affiliation{Department of Astronomy, University of Illinois at Urbana-Champaign, 1002 West Green Street, \\
Urbana,IL 61801, U.S.A.}
\affiliation{National Center for Supercomputing Applications, 1205 West Clark St., Urbana, IL 61801, U.S.A.}

%
%

%
%
%
\begin{abstract}

The Large Magellanic Cloud (LMC) has $\sim$60 confirmed supernova remnants (SNRs).  Because of the known distance, 50 kpc, the SNRs' angular sizes can be converted to linear sizes, and their X-ray observations can be used to assess X-ray luminosities ($L_X$).  We have critically examined the LMC SNRs' sizes reported in the literature to determine the most plausible sizes.  These sizes and the $L_X$ determined from \emph{XMM-Newton} observations are used to investigate their relationship in order to explore the environmental and evolutionary effects on the X-ray properties of SNRs.
We find: (1) Small LMC SNRs, a few to 10 pc in size, are all of Type Ia with $L_X>10^{36}$ ergs s$^{-1}$. The scarcity of small core-collapse (CC) SNRs is a result of CCSNe exploding in the low-density interiors of interstellar bubbles blown by their massive progenitors during their main sequence phase. (2) Medium-sized (10-30 pc) CC SNRs show bifurcation in $L_X$, with the X-ray-bright SNRs either in an environment associated with molecular clouds or containing pulsars and pulsar wind nebulae and the X-ray-faint SNRs being located in low-density interstellar environments. (3) Large (size$>$30 pc) SNRs show a trend of $L_X$ fading with size, although the scatter is large. The observed relationship between $L_X$ and sizes can help constrain models of SNR evolution.
\end{abstract}
\subjectheadings{
ISM: supernova remnants --- supernovae: general --- X-rays: ISM --- Magellanic Clouds
}
%
%

\section{Introduction}  \label{sec:intro}

Most supernova remnants (SNRs), regardless of progenitor types, exhibit some kind of X-ray emission. Thermal emission can arise from shocked interstellar medium (ISM) and/or SN ejecta, while relativisic electrons interacting with amplified magnetic field can produce non-thermal (synchrotron) emission.  In the cases of core-collapse (CC) SNRs, there may exist additional X-ray emission from pulsars and pulsar wind nebulae (PWNe).  See \citet{vink2012} for a comprehensive review of X-ray emission from SNRs.

To make a statistical study of X-ray emission of SNRs, we need a large sample of SNRs with known distances.  The Galactic sample of SNRs is quite incomplete because of heavy absorption in the Galactic plane, and the distances to individual SNRs are often very uncertain.  The Large Magellanic Cloud (LMC), on the other hand, has small internal and foreground absorption column densities \citep{schlegel1998}, and hosts a large sample of SNRs all at essentially the same known distance 50 kpc\footnote{Note that due to the LMC's inclination of 18-23 degrees in the line of sight, the error in the distance and linear size can be uncertain by up to 10\% \citep{subramanian2010}, and the luminosity can be uncertain by 20\%. These uncertainties, however, do not affect the general conclusions of these paper.} \citep{pietrzynski2013}. At least 59 SNRs have been confirmed and a few additional SNR candidates have been suggested \citep{maggi2016,bozzetto2017}.  This large sample of LMC SNRs is ideal for systematic and statistical studies of X-ray emission from SNRs.

Recently, \citet{maggi2016} analyzed {\it XMM-Newton} observations of the 59 confirmed SNRs in the LMC, deriving physical properties of the X-ray-emitting plasma from spectral fits.  Because of the known distance, it is possible to determine the X-ray luminosity of each SNR.  In the meantime, \citet[][hereafter Bo2017]{bozzetto2017} measured the sizes of the 59 LMC SNRs using X-ray, radio and optical images.  Intrigued by these results, we have examined the relationship between X-ray luminosity and size of LMC SNRs in order to explore evolutionary effects and environmental impacts on X-ray  properties of SNRs.

This paper reports our investigation of the relationship between X-ray luminosity and size of SNRs in the 
LMC.  In Section 2, we discuss the physical meaning of SNR sizes measured at optical or X-ray wavelengths,
examine the SNR sizes reported in the literature, and assess the most reliable sizes that represent the SNRs' full
extent.  In Section 3, we plot X-ray luminosities against sizes for LMC SNRs and note intriguing features in the distribution
of SNRs in this plot.  In Section 4, we discuss the physical reasons behind the distribution of SNRs in the plot of
X-ray luminosity versus size.  Finally, a summary is given in Section 5.

\begin{figure*}[tbh]
\centering
\includegraphics[scale=0.35]{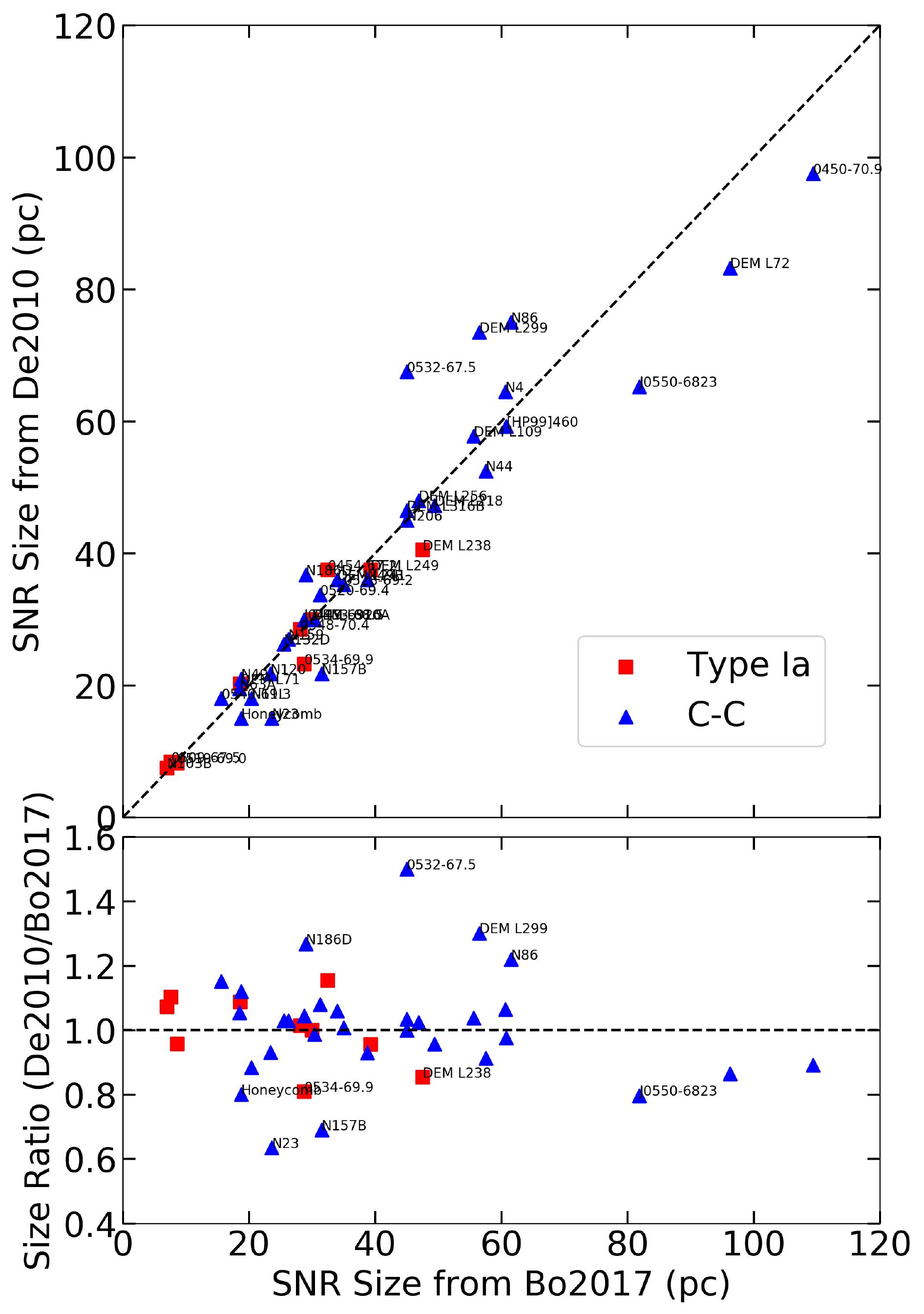}
\includegraphics[scale=0.35]{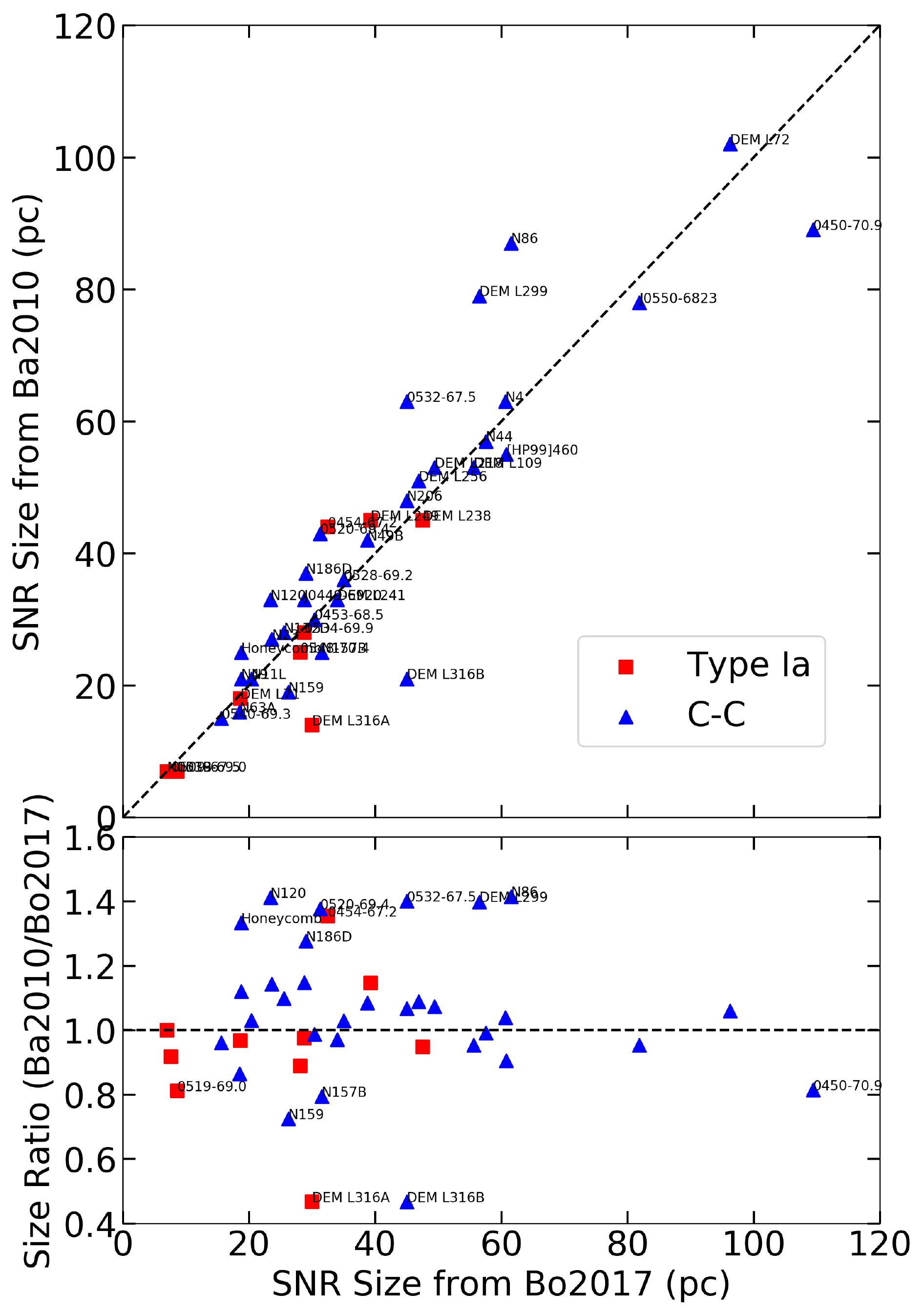}
\caption{The left panels compare the SNR sizes reported by De2010 and Bo2017, with the upper panel 
plotting De2010 sizes versus Bo2017 sizes and the lower panel plotting the (De2010 size / Bo2017 size) ratios
versus Bo2017 sizes.   The right panels compare the SNR sizes reported by Ba2010 and Bo2017 in the same way
as the left panels.
}
\end{figure*}

\section{Sizes of LMC SNRs}  \label{sec:LxSize}
Both X-ray and optical images have been used to measure SNR sizes, but it should be noted that while X-ray and optical H$\alpha$ emission both originate from post-shock gas, they arise under different physical conditions.
Generally speaking, X-ray emission comes from hot gas with temperatures $\gtrsim 10^6$K, 
while H$\alpha$ emission originates from ionized gas at $\sim 10^4$K; therefore, an SNR size measured
in X-rays may differ from that measured in H$\alpha$.  

Measurements of X-ray and H$\alpha$ sizes of SNRs can also differ because of different instrumental sensitivities. For example, the \emph{XMM-Newton} observations of the LMC SNRs detect volume emission measures ($EM_V \equiv \int n_en_HdV$, where $n_e$ is the electron density, $n_H$ is the hydrogen density, and $V$ is the emitting volume) of 10$^{54}$ -- 10$^{60}$ cm$^{-3}$ \citep{maggi2016}. For a spherical volume of highly ionized interstellar gas (${n_e}/{n_H} \sim 1.2$ for a typical helium to hydrogen number ratio of 0.1), the rms density derived from the volume emission measure is
\begin{equation}
\langle n_H^2 \rangle^{1/2}=\bigg(\frac{5~EM_V}{ \pi f D^3}\bigg)^{1/2},
\end{equation}
where $D$ is the diameter of the X-ray emitting gas, and $f$ is the volume filling factor. For SN ejecta dominated by heavy elements, ${n_e}/{n_H}$ should be greater than 1.2, and hence the rms hydrogen density in equation (1) is the upper limit, which is about  (0.001--3) $f^{-1/2}$ cm$^{-3}$ for the LMC SNRs as derived from the \emph{XMM-Newton} measurements.  Meanwhile, narrow-band H$\alpha$ CCD images can easily detect emission measures ($EM_{\ell} \equiv \int n_en_Hd\ell$, where $\ell$ is the emitting path length) down to 10--20 cm$^{-6}$ pc, and for an emitting path length of 5 pc the electron density needs to be at least 1.4--2 cm$^{-3}$.  Thus, for SNRs running into a dense medium with
densities $>$ 1 H-atom cm$^{-3}$, the shocked gas can be detected in both X-ray and optical wavelengths, while 
those running into a medium with densities $\ll$1 H-atom cm$^{-3}$ may be visible in X-rays but not in optical.

Another factor that can affect the measurements of SNR sizes is the wavelength-dependent confusion from the background.  SNRs may be located adjacent to HII regions or superposed on a complex background, in which case the boundary of an SNR can be diagnosed by sharp filamentary morphology, enhanced [\ion{S}{2}] line emission, high-velocity components in optical emission lines, nonthermal radio emission, and diffuse X-ray emission \citep{chu1997}.  When more than one of the above diagnostics are detected, the SNR boundary can be more reliably measured.  However, if X-ray emission is the only diagnostic detected and the SNR emission is superposed on a large-scale diffuse X-ray emission, the background confusion can prevent accurate measurements of the SNR size.   

Several publications have reported sizes of the LMC SNRs, but there are often discrepancies between their measurements. \citet[][hereafter Ba2010]{badenes2010} used mainly X-ray images from \emph{Chandra} or \emph{XMM-Newton} to determine the SNR sizes, and adopted previous optical and radio measurements when high-resolution X-ray images were not available. \citet[][hereafter De2010]{desai2010} considered optical and X-ray images, and measured SNR sizes based on the extent of diffuse X-ray emission or filamentary H$\alpha$ shell structure. Bo2017 considered optical, X-ray and radio images. 
\citet{maggi2016} also listed SNR sizes, but they only gave the maximal diameters in X-ray; thus these sizes are often much larger than the ones reported by the above three references.  Below we compare the SNR sizes reported by Ba2010, De2010, and Bo2017.

Bo2017 has the largest and most complete SNR sample, and is thus chosen to be the reference for comparisons.
Figure 1 compares SNR sizes reported by De2010 and Bo2017 in the left panels, and Ba2010 and Bo2017 in the right panels.  The upper panels plot SNR sizes from one source versus another, while the lower panels plot the ratios of SNR sizes from two sources.  

De2010 and Bo2017 both used primarily optical and X-ray images for the SNR size measurements, but there are still discrepancies greater than 16\% and up to 50\%. The discrepancies are caused by the following reasons: (1)  
The SNR sizes can be measured only in X-rays and the surface brightness varies significantly, such as N23,
or the background is complex, such as the Honeycomb and 0532-67.5; in such cases the discrepancy in size measurements can be as high as 50\%.  (2) The SNR is superposed on an \ion{H}{2} region or a superbubble, whose H$\alpha$ emission can confuse the size measurements, such as N157B and N186D.  (3) The SNR size is measured without simultaneously considering optical, X-ray, and radio images that show wavelength-dependent distribution of emission, such as 0534$-$69.9, DEM\,L238, DEM\,L299, and J0550$-$6823.  (4) The irregular shape of an SNR can cause subjective size measurements to differ by up to $\sim$20\%, such as N86. For these discrepant objects, we examine their H$\alpha$, [\ion{S}{2}], X-ray, and 24 $\mu$m images (in Appendix A), consider radio and kinematic properties available in the literature, and make new measurements (described in Appendix B).  

The comparisons between Ba2010 and Bo2017 sizes, right panels of Figure 1, show numerous large discrepancies.  These discrepancies are caused by the larger uncertainties in Ba2010 sizes that were compiled from previous measurements based on mainly X-ray images and some optical images.  As mentioned above and detailed in Appendices A--B, multi-wavelength examination of an SNR provides the most comprehensive picture of its physical structure and boundaries, and size measurements based on only one single wavelength may not reflect the SNR's true extent.

\begin{figure*}[tbh]
\centering
\includegraphics[scale=0.5]{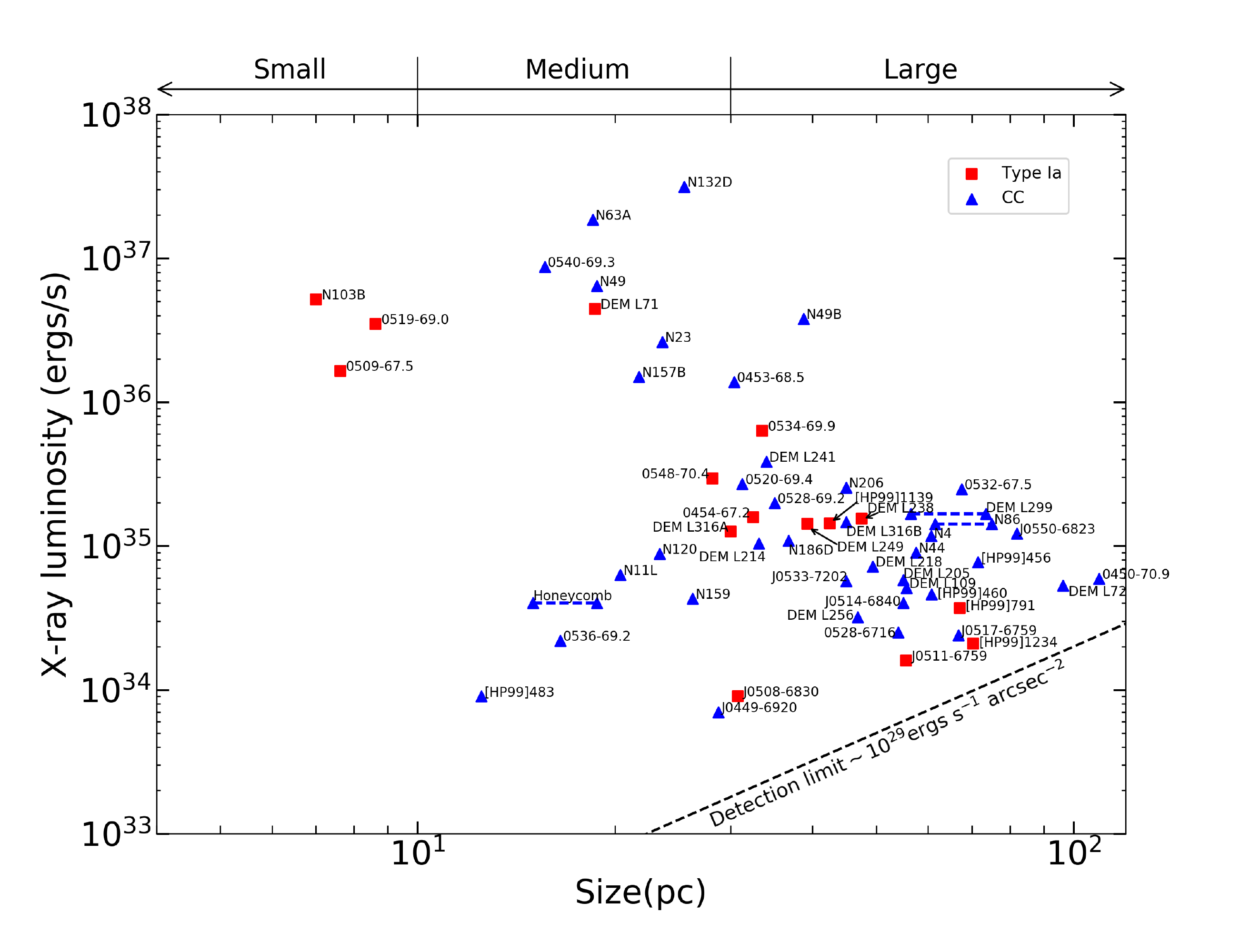}
\caption{Observed $L_X$ versus size plot for SNRs in the LMC.  The size and $L_X$ are listed in Appendix C. 
For the Honeycomb SNR and N86, we have used
two extreme size measurements to illustrate the
largest uncertainties in the size measurements.
Note that some of the large CC SNRs may in fact be Type Ia SNRs 
whose SN ejecta have cooled and no longer emit detectable X-rays 
for them to be identified as such.
}
\end{figure*}
\begin{figure*}[tbh]
\centering
\includegraphics[scale=0.5]{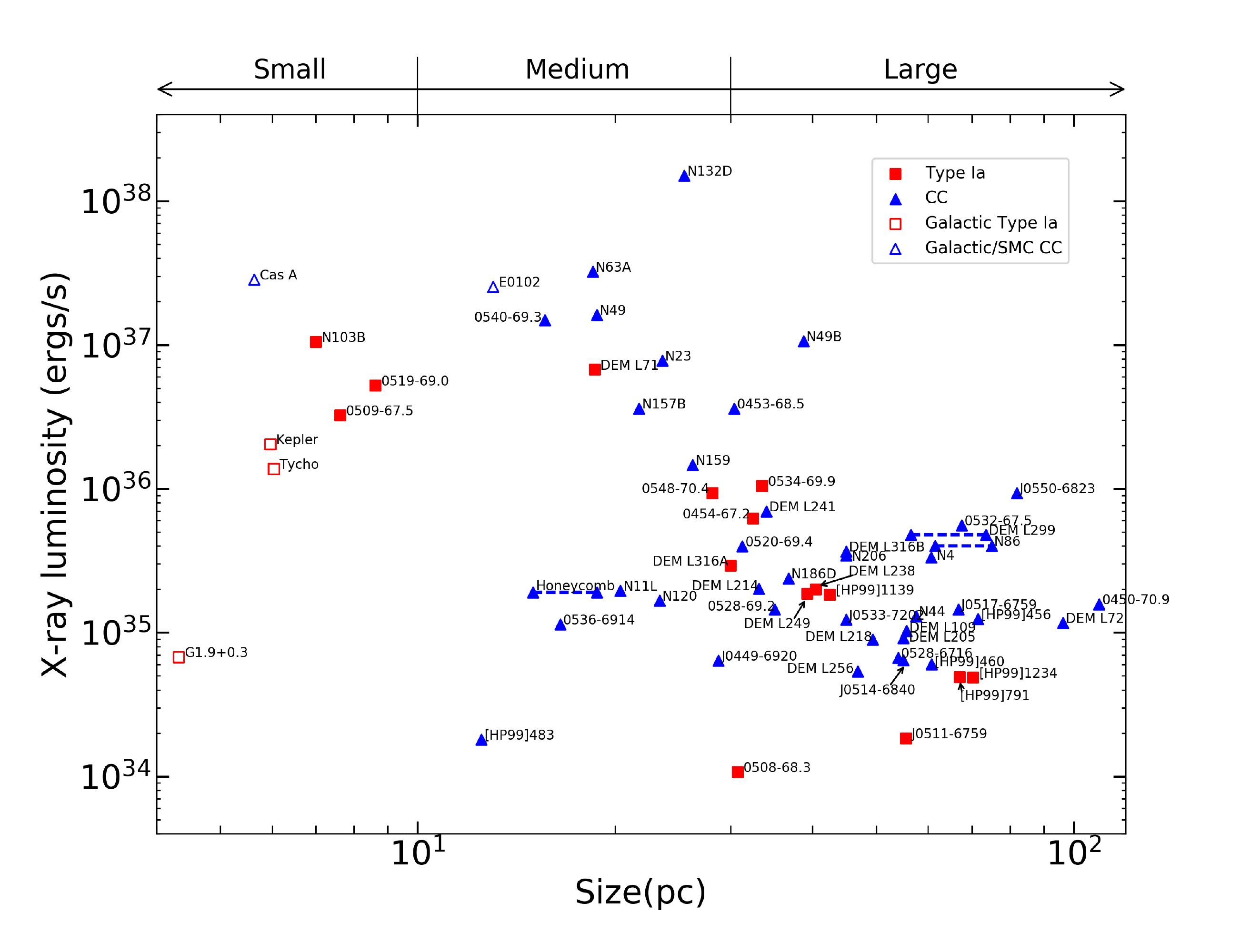}
\caption{Unabsorbed $L_X$ versus size plot for SNRs in the LMC.
}
\end{figure*}
\section{Relationship between sizes and $L_X$}  \label{sec:LxSize}
The LMC provides an ideal sample of SNRs for us to study the relationship between their
X-ray luminosities ($L_X$) and sizes.  The size of an SNR may be intuitively thought to reflect its 
evolutionary stage, because an SNR expands as the shock wave propagates outward and a large SNR would be older than a small SNR, if the ambient interstellar densities are similar.  However, the ambient 
ISM does have a wide variety of physical properties and conditions, and the relationship between size
and evolution can be quite complex.  Through
the relationship between $L_X$ and size we hope to investigate effects of ambient environment
and evolution on the SNRs' X-ray luminosities.

SNRs rarely show round, symmetric shell structures with well-defined sizes.
To assign a ``size'' to an irregular SNR, we adopt the average of its major and 
minor axes.  Such SNR sizes determined from data of Ba2010, De2010, and Bo2017 
are tabulated in Appendix C.   As discussed at length in Section 2, De2010 and Bo2017 
sizes were determined primarily with optical and X-ray images of SNRs and are in agreement
for most cases.  
For 83\% of the SNRs that have Bo2017 and De2010 sizes differ by less than 16\%, we adopt their average sizes from Bo2017.
For the SNRs with larger discrepancies between De2010 and Bo2017, we
discuss individual objects and determine their average sizes in Appendix B, and list their
adopted sizes in the table in Appendix C.

The LMC sample of SNRs have been studied in X-rays with \emph{XMM-Newton} by 
\citet{maggi2016}.  They fit the X-ray spectra with the package XSPEC \citep{arnaud1996} 
using a combination of collisional ionization equilibrium (CIE) models and non-equilibrium 
ionization (NEI) models, derived their X-ray fluxes in the 0.3 to 8 keV band, and computed 
their $L_X$ for an LMC distance of 50 kpc  \citep{pietrzynski2013}.  These observed (absorbed) $L_X$ 
are listed in the last column of the Table in Appendix C.

Using the $L_X$ and average sizes in Appendix C, we plot the $L_X$ versus size 
for the LMC SNRs in Figure 2. We have also made the same $L_X$--size plot with unabsorbed $L_X$ and present it in Figure 3.  (These unabsorbed $L_X$ are from the same model fits that produced the absorbed $L_X$ published by \citealt{maggi2016}).  The distribution of the SNRs are qualitatively similar to that in Figure 2. 
Note that we did not include SN 1987A (size$=0.45$ pc, $L_X=2.7\times 10^{36}$ ergs s$^{-1}$) because its inclusion will leave vast empty space on the left and compress all the data points on the right in Figure 2 and 3.

At first glance, the $L_X$ -- size plot for LMC SNRs shows a scattered diagram; however, if the sizes are divided
into three ranges: $<$10 pc as ``small'', 10--30 pc as ``medium'', and $>$30 pc as ``large'', it is possible
to see interesting trends in each size range: \\
(1) For sizes a few to 10 pc, only a small group of SNRs exist with $L_X$ of a few $\times$ 
10$^{36}$ ergs s$^{-1}$, and all of them are of Type Ia.  For comparison, we add the Galactic SNRs 
with sizes a few to 10 pc in Figure 3; the data are taken from the Chandra Supernova Remnant Catalog\footnote{See http://hea-www.cfa.harvard.edu/ChandraSNR/.}.
 Interestingly, Tycho and Kepler SNRs,
two small Type Ia SNRs in our Galaxy, are also located in the similar part of $L_X$--size plot
as the small LMC Type Ia SNRs. 
In contrast, the Galactic CC SNR Cas A is an order of magnitude more luminous than these small Type Ia SNRs, and the $\sim$100-year old Galactic Type Ia SNR G1.9+0.3 is  smaller and significantly fainter \citep{reynolds2008,borkowski2010,borkowski2013}.\\
(2) For sizes 10--30 pc, there is a bifurcation in the distribution of SNRs.  The X-ray-bright ones have 
$L_X > 10^{36}$ ergs s$^{-1}$ and the X-ray-faint ones have $L_X < 10^{35}$ ergs s$^{-1}$ for sizes
below $\sim$20 pc, and these two groups converge to a few $\times$10$^{35}$ ergs s$^{-1}$ towards 30 pc size.
It is worth noting that the X-ray-faint medium-sized SNRs are mostly CC SNRs.\\
(3) For sizes $>$30 pc, while $L_X$ exhibits a wide range, the majority of the SNRs appear to show a general trend of $L_X$ decreasing with size.

It also appears that the Type Ia SNRs show smaller scatter 
in $L_X$ for any given size than the CC SNRs, especially in the medium 
size range (Figures 2 and 3).  The scatter in $L_X$ reflects the ambient 
interstellar density: Type Ia SNe occur in diffuse medium with moderate 
densities, while CC SNe can take place near dense molecular clouds or 
in a very low-density environment produced by energy feedback from 
massive stars.  Because of the smaller scatter in $L_X$, the smooth
variations of Type Ia SNRs' $L_X$ versus size may demonstrate the
SNR evolution.

The dashed line in the lower right corner of Figure 2 corresponds to a constant
surface brightness of $10^{29}$ ergs s$^{-1}$ arcsec$^{-2}$, which represents the typical detection limit of the \emph{XMM-Newton} observations used by \citet{maggi2016}.
Consequently, no SNRs are located beneath this dashed line.

\section{Discussion}  \label{sec:Discussion}

We have examined the physical structures and environments of SNRs in the three size ranges 
in order to understand the physical significance of their distributions in the $L_X$--size plot.
The discussion in this section is ordered according to the SNR sizes.

\subsection{Small Known LMC SNRs Are Dominated by Type Ia}
It is striking that the small LMC SNRs, with sizes a few to 10 pc, are all Type Ia SNRs with
$L_X$ of a few $\times$ 10$^{36}$ ergs s$^{-1}$. 
(Note that SN 1987A is outside the size range under discussion.) 
For comparison, we show that the Galactic Type Ia SNRs Kepler and Tycho are both located 
in a similar region as the 
 young LMC Type Ia SNRs. 
The small range of $L_X$ for small Type Ia SNRs and the scarcity
  of small CC SNRs can be explained as follows. 
  
Type Ia SNe are usually considered to explode in a tenuous and uniform ISM \citep[e.g.,][]{badenes2005}. 
On the other hand, CC SNe usually explode inside interstellar bubbles blown by the fast stellar winds 
of their massive progenitors during the main sequence phase  \citep{castor1975,weaver1977}.  Interstellar
bubble interiors have very low densities, and hence CC SNe inside bubbles are called ``cavity explosions''.
It is conceivable that the interstellar environments of Type Ia and CC SNe have very different density profiles.

Density profiles of ambient medium strongly affect the evolution of an SNR's $L_X$.
In a classical model of a SN explosion in a 
uniform ISM, the resulting SNR goes through free expansion phase, Sedov 
phase (i.e., adiabatic phase), and radiative phase \citep{woltjer1972}.  
The Sedov phase starts when the swept-up ISM mass is several times the
SN ejecta mass \citep[e.g.,][]{dwarkadas1998}.  The $L_X$ of an SNR
during the Sedov phase can be calculated \citep[e.g.,][]{hamilton1983}.
To illustrate the evolution of $L_X$ for different ambient densities, we plot
$L_X$ against age and size in Figure 4.

\begin{figure*}[tbh]
\centering
\includegraphics[scale=0.3]{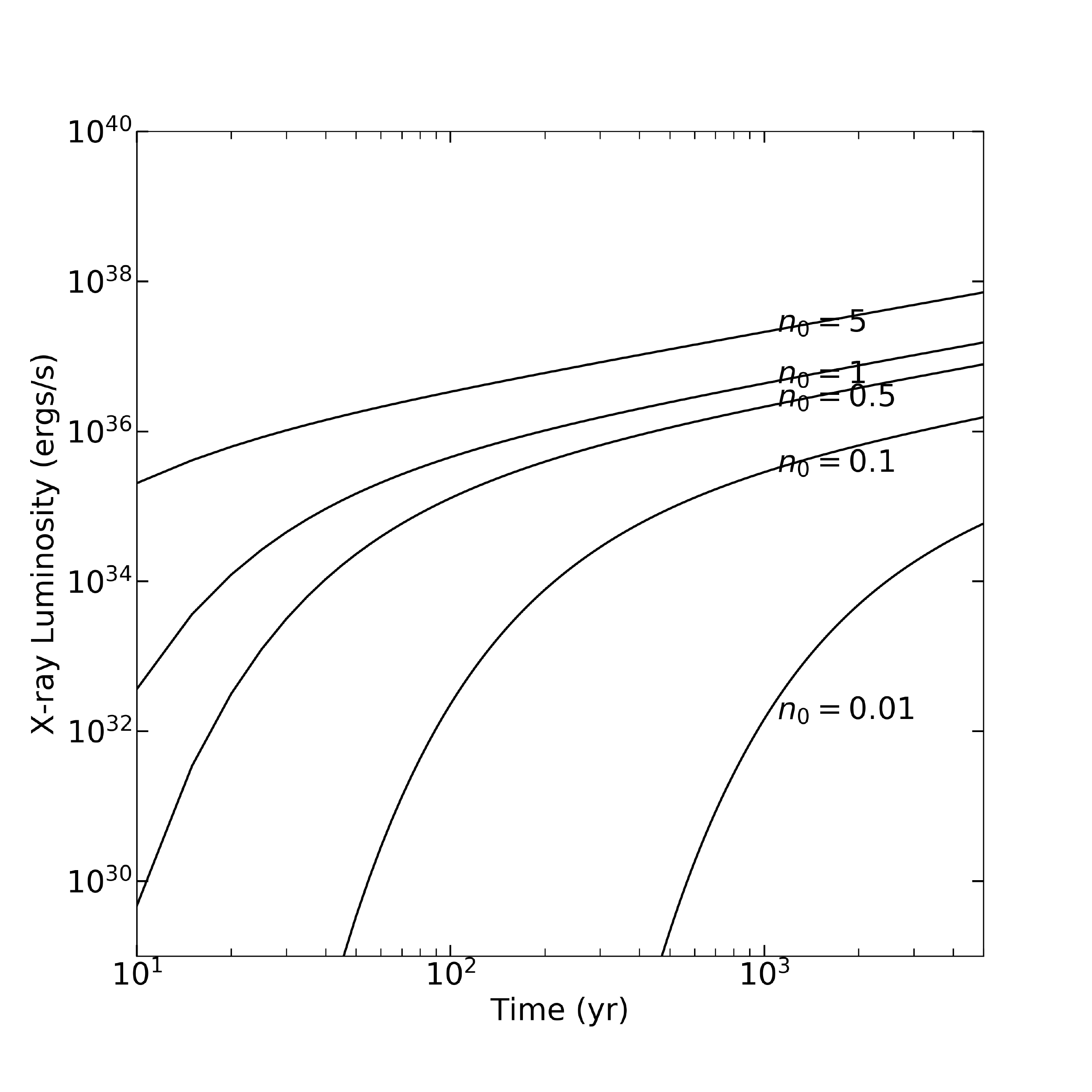}
\includegraphics[scale=0.3]{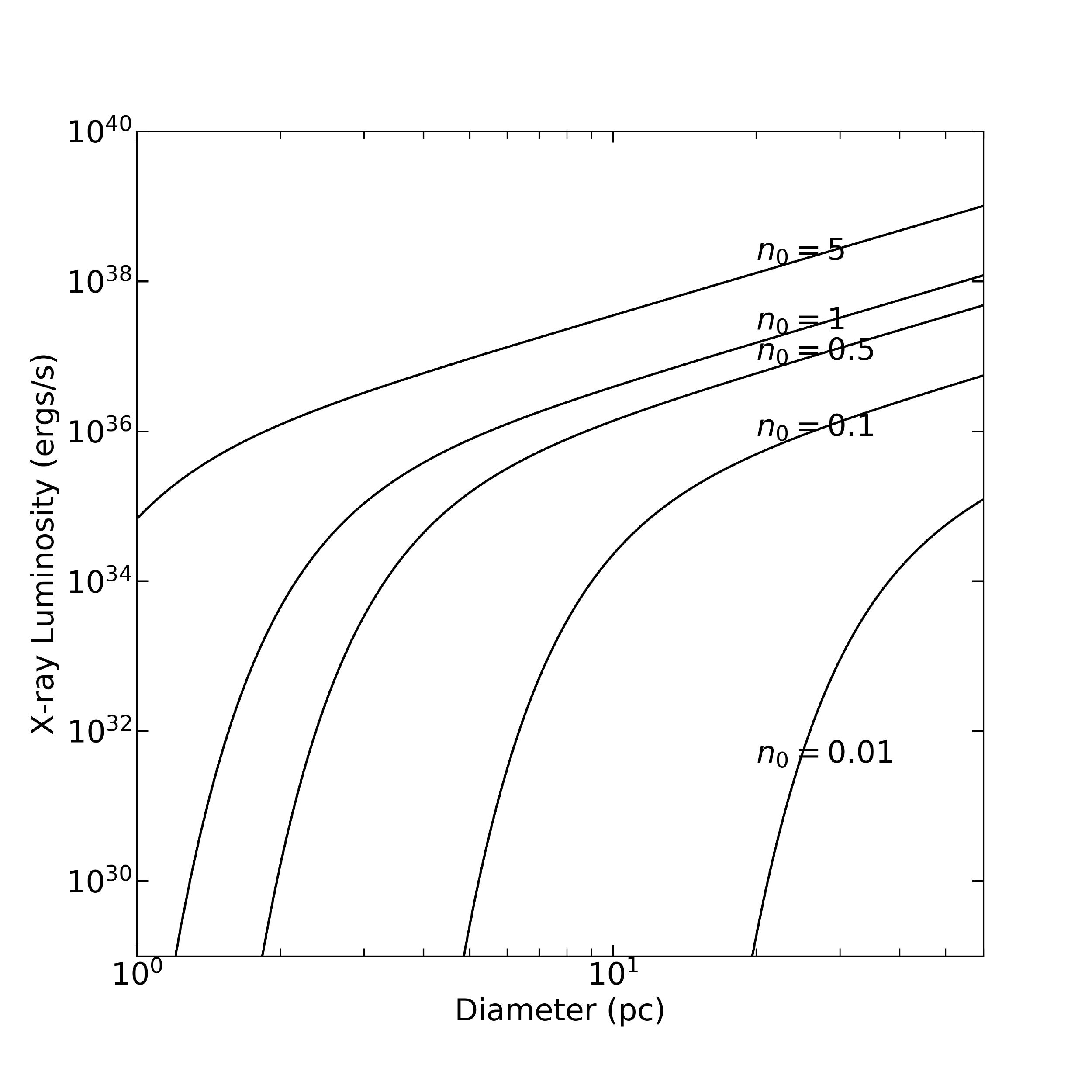}
\caption{Evolution of $L_X$ in the Sedov model. The ambient density $n_0$ in H-atom cm$^{-3}$ is marked for each model.
}
\end{figure*}

For a Type Ia SNR in a partially neutral ISM, only the ionized interstellar gas can be swept up by the shock.  Thus, for a uniform density of $\sim$1 H-atom cm$^{-3}$ and a neutral fraction of $\eta$, the Sedov phase will start when the swept-up ionized gas reaches 1.4 $M_\odot$ in mass, corresponding to a radius of 2.4$(1-\eta)^{-1/3}$ pc.  This radius is 5.2 pc if $\eta$ = 0.9, and 3 pc if $\eta$ = 0.5.  These sizes are comparable to the young Balmer-dominated Type Ia SNRs in the LMC, 0509$-$67.5 and 0519$-$69.0; thus, it is likely that these young Type Ia SNRs are entering the Sedov phase.  However, the interstellar density is so much lower than the SN ejecta density that their X-ray emission is still dominated by that produced by the reverse shock into the SN ejecta.  This is evidenced in the SN ejecta abundance revealed in the X-ray spectra of these small Balmer-dominated Type Ia SNRs, although the X-ray emission shows a shell morphology \citep{warren2004,kosenko2010}.  
The larger Type Ia SNRs, such as DEM\,L71 and 0548$-$70.4 with sizes in the 20-30 pc range, must be in the Sedov phase already.  Furthermore, their forward shock and reverse shock have traveled farther apart, and their X-ray emission shows the forward shock in an interstellar shell well resolved from the reverse shock in the SN ejecta \citep{hughes2003,hendrick2003}.

X-ray emission from reverse shocks is the cause of the high $L_X$ of small Type Ia SNRs.
The small scatter of these young bright Type Ia SNRs in the $L_X$ vs size plot
reflects their similar ages, the relative uniformity of SNe Ia (in term of nucleosynthesis and explosion energy), and the modest effect the progenitors have on changing their immediate surrounding. 
The smallest Galactic Type Ia SNR G1.9+0.3 has a low $L_X$ because it is so young ($<$200 yr) that the reverse shock has only gone through very little of the SN ejecta \citep{reynolds2008,borkowski2014}.

For CC SNRs whose SNe exploded in cavities of wind-blown bubbles, due to the 
extremely low density within the bubbles ($\sim 10^{-4}-10^{-2}$ H-atom cm$^{-3}$), 
the X-ray emission from shocked gas would be too faint to be detected at a young 
age; only when the SNR's forward shock hits the dense shell/wall of a bubble will 
the X-ray luminosity jump up several orders of magnitude \citep{dwarkadas2005}.  
As shown by \citet{naze2001}, main sequence O stars have interstellar bubbles of 
sizes 15--20 pc.  By the time a massive star explodes as a CC SN, its main-sequence 
bubble has grown larger, and hence the SNR shock goes through the low-density bubble
interior without producing detectable X-ray emission until it hits the bubble shell wall at 
radius of 10 pc or larger. 

For illustration, considering a spherical interstellar bubble with a radius of 10 pc and 
assuming a simplistic extreme case (upper limit) of average density of 0.01 H-atom cm$^{-3}$ in 
the bubble interior, we can calculate the total mass in the bubble interior to be 
$\lesssim$ 1 $M_\odot$; thus, when the SNR shock reaches the bubble wall, it has swept 
up only $\sim$1 $M_\odot$, much lower than the CCSN ejecta mass, a few to a few tens 
$M_\odot$; thus, the Sedov phase has not been reached.  The bubble shell 
consists of swept-up ISM that was originally distributed in the bubble cavity.  Assuming 
the bubble was blown in a diffuse ISM with density of 1 H-atom cm$^{-3}$, the 
total mass in the bubble shell would be 100 $M_\odot$; therefore, the SNR
reaches the Sedov phase when the forward SNR shock traverses the bubble shell.

During the free-expansion phase, the SNR shock is not significantly decelerated and it 
remains fast until it hits the bubble wall.  Assuming a constant shock velocity of 
10,000 km s$^{-1}$, it only takes 1000 years for the SNR to grow to a radius of 10 pc.
Consequently, SNRs inside interstellar bubbles not only emit very faintly in X-rays, but 
also expand very rapidly to reach the dense shell wall.  Such ``cavity explosions'' explain 
the absence of small CC SNRs in the $L_X$--size plot. 
Cavity explosions are also responsible for the discrepancies between ionization ages and dynamical
ages of LMC SNRs, such as N132D, N63A, and N49B \citep{hughes1998}.

We have plotted the young CC SNR Cas A in Figure 3 for comparison.
Cas A is small in size and luminous in X-rays.  These properties are
caused by its interaction with a dense circumstellar medium, i.e.,
material ejected by the SN progenitor \citep{fesen2001}.  Circumstellar 
bubbles are often observed around Wolf-Rayet stars and luminous 
blue variables (LBVs), and circumstellar bubbles are smaller than
interstellar bubbles \citep{chu2003}.  Cas A SN must have exploded in a circumstellar bubble.

\subsection{X-ray-Bright and X-ray-Faint Medium-Sized SNRs}

The medium-sized LMC SNRs show clear bifurcation in their $L_X$.  In the X-ray-bright group 
with $L_X \ge 10^{36}$ ergs s$^{-1}$, only one is of Type Ia, and the other seven are 
CC SNRs.  Among these X-ray-bright CC SNRs, four are interacting with molecular clouds, 
as CO emission was detected near the SNRs N23, N49, and N132D \citep{banas1997,
park2003} and H$_2$ absorption is detected in \emph{Spitzer} IRS observations
towards N63A (Segura-Cox et al.\ 2018, in preparation).  None of these four X-ray-bright CC SNRs
show sharp H$\alpha$ shell structure enclosing the diffuse X-ray emission, indicating that the forward
SNR shocks are still in the low-density interiors of bubbles.  
In the cases of N23 and N132D, where no prominent shocked 
cloudlets are seen,  the X-ray emission does show limb-brightening, 
indicating that the ambient medium is dense enough to produce 
detectable X-ray emission but not optical H$\alpha$ emission, and 
this ambient medium may correspond to the conduction layer in a
bubble interior \citep{weaver1977}.  As N23 and N132D are both 
associated with molecular clouds, their bubble shells and conduction
layers must have higher densities, which contribute to the bright X-ray 
emission.
In the cases of N49 \citep{bilikova2007, park2012} and N63A \citep{warren2003}, it is clear that dense cloudlets,
possibly associated with the molecular clouds, have been shocked
and contribute to the X-ray emission. 
 The other three X-ray-bright CC SNRs possess 
bright PWNe: 0540$-$69.3 \citep{gotthelf2000}, N157B \citep{wang1998}, and 
0453$-$68.5 \citep{gaensler2003}.  
Pulsars and PWNe are powerful sources of nonthermal X-ray emission
and provide additional X-ray emission to boost their SNRs' total $L_X$.
Note that the PWN of 0453$-$68.5 is not particularly dominating, but its X-ray image show a
limb-brightened sharp shell that indicates
that the shock has already reached the bubble shell. While 0453$-$68.5 has a PWN, it is the SNR
shock impact on the dense bubble shell giving rise to $L_X$.

The X-ray-faint medium-sized SNRs are mostly associated with CC SNe.  Among the three X-ray-faint CC SNRs
smaller than 20 pc, 0536$-$69.2 and [HP99]483 are not detected in optical, and the Honeycomb SNR
shows only a small patch of honeycomb-like nebulosity resulting from SNR shocking a piece of 
shell wall \citep{chu1995, meaburn2010}.  The absence of sharp optical shells enclosing the diffuse 
X-ray emission indicates a low-density ISM around these SNRs.  The Honeycomb SNR has hit a small
piece of dense gas and hence it has the highest $L_X$ among these three, but still a couple orders of
magnitude fainter than the SNRs interacting with molecular clouds.  The X-ray-faint SNRs with sizes
20--30 pc all show optical shell structure enclosing their diffuse X-ray emission, and they have higher $L_X$
than the smaller ones, except J0449$-$6920, whose \emph{XMM-Newton} observation was too shallow to make
accurate measurements.  These CC SNRs may represent cavity explosions whose SNR shocks have
just reached the bubble shell walls. The SNRs N11L and N120 have just reached the bubble shell, but the bubble shell densities are not as high as those of N23 and N132D.

\subsection{Fading of X-rays in Large SNRs}

Among the large (size $>$30 pc) LMC SNRs, a general trend of decreasing $L_X$ for larger SNRs can be seen, but for any given size, the differences in $L_X$ can be up to one order of magnitude.

As an SNR sweeps up more interstellar gas, the shock velocity decreases and when it goes much below $\sim$300 km s$^{-1}$, the post-shock temperature will be below 10$^6$ K, too low to generate X-ray-emitting gas.  The hot gas in SNR interior cools, and the X-ray emission diminishes.

The scatter in $L_X$ may be caused by the differences in ambient gas densities ($n_0$) and the SN explosion energies ($E$). To evaluate the effects of these two factors, we consider a spherical SNR of radius $R$, whose X-ray emission originates from shocked ISM in a shell.  Its $L_X$ is $\propto$ (emitting volume) $\times$ (density)$^2$ $\times$ (emissivity).  As (emitting volume) $\times$ (density) is proportional to the total interstellar mass within radius $R$, it is $\propto$ $R^3n_0^2$.  The emissivity is a slow function of temperature for photon energies below 2 keV \citep{hamilton1983}.  Since the large old SNRs are likely at low X-ray emitting temperatures, a few $\times$10$^6$ K at most, we will treat the emissivity as a constant, and $L_X$ $\propto$ $R^3n_0^{2}$.

The total kinetic energy in the SNR shell scales with the explosion energy, so $E \propto R^3n_0 v^2$.  The large old SNRs have low expansion velocities of a few $\times 10^2$ km s$^{-1}$, so we will also approximate the expansion velocity as a constant.  Thus, $L_X$ $\propto$ $E n_0$ \footnote{ Note that this is in interesting contrast with the radio luminosity, which scales as $L_{radio}$ $\propto$ $E^{1.3} n_0^{0.45}$ or $L_{radio}$ $\propto$ $E^{1.45} n_0^{0.3}$ depending on the magnetic field amplification mechanism by the shock \citep{chomiuk2009}.}.  The effects of the ambient density and the SN explosion energy are about equally important. However, the ranges of the ambient gas densities and the SN explosion energies are quite different. The ambient interstellar density can range from 0.01 to a few hundred H-atom cm$^{-3}$, about 4 orders of magnitude, while the SN explosion energies are mostly clustered around 10$^{51}$ ergs with extreme values differing by no more than 3 orders of magnitude \citep[e.g.,][]{woosley1986}.

Hence, the large scatter in $L_X$ for SNRs with the same size is more likely caused by the detailed differences in the ambient gas densities, and the SN explosion energy plays a lesser role in raising the scatter in $L_X$.

\section{Summary}

The LMC is at a known distance of 50 kpc, and thus the linear sizes of SNRs in the LMC
can be determined from their angular size measurements \citep{desai2010,bozzetto2017}, 
and their X-ray luminosities can be determined from \emph{XMM-Newton} X-ray observations 
\citep{maggi2016}, allowing a unique opportunity to examine the relationship 
between $L_X$ and size of SNRs.  We have critically compared LMC SNR sizes
reported by different authors in the literature and adopted the most reasonable sizes to 
investigate how $L_X$ vary with sizes among the LMC sample of SNRs.

We find that the $L_X$ -- size relationship for LMC SNRs can be divided into small, medium,
and large size ranges:\\
(1) The small LMC SNRs with sizes a few to 10 pc are all young Type Ia SNRs with $L_X$ a few
times 10$^{36}$ ergs s$^{-1}$.  The apparent scarcity of small CC SNRs may be caused by
their ``cavity explosions'', as massive progenitors of CCSNe have blown interstellar bubbles
and the SN explosions take place in the very low-density interiors of the bubbles. \\
(2) The medium-sized SNRs, with sizes 10--30 pc, show bifurcation in their $L_X$ with
an order of magnitude difference in $L_X$.  The X-ray-bright CC SNRs either are 
in an environment associated with molecular clouds or have pulsars and pulsar-wind nebulae emitting
nonthermal X-ray emission.  \\
(3) The large SNRs, with sizes greater than $\sim$30 pc, show a general trend of
fading $L_X$ at large sizes.  As these sizes are larger than the normal interstellar 
bubbles blown by massive stars, the large SNRs have swept up the bubble material
and extended into the diffuse ISM.  As the SNR shocks sweep up more ISM, the shock
velocity slows down.  When the post-shock velocities are too low to produce X-ray-emitting
material, the hot plasma in SNR interiors cool and reduce the X-ray emission.

\acknowledgments
This project is supported by Taiwanese Ministry of Science and 
Technology grant MOST 104-2112-M-001-044-MY3.

\appendix
\section{Images of the SNRs with Large Discrepancies between different Size Measurements}
\begin{figure*}[h]
\centering
\includegraphics[scale=0.25]{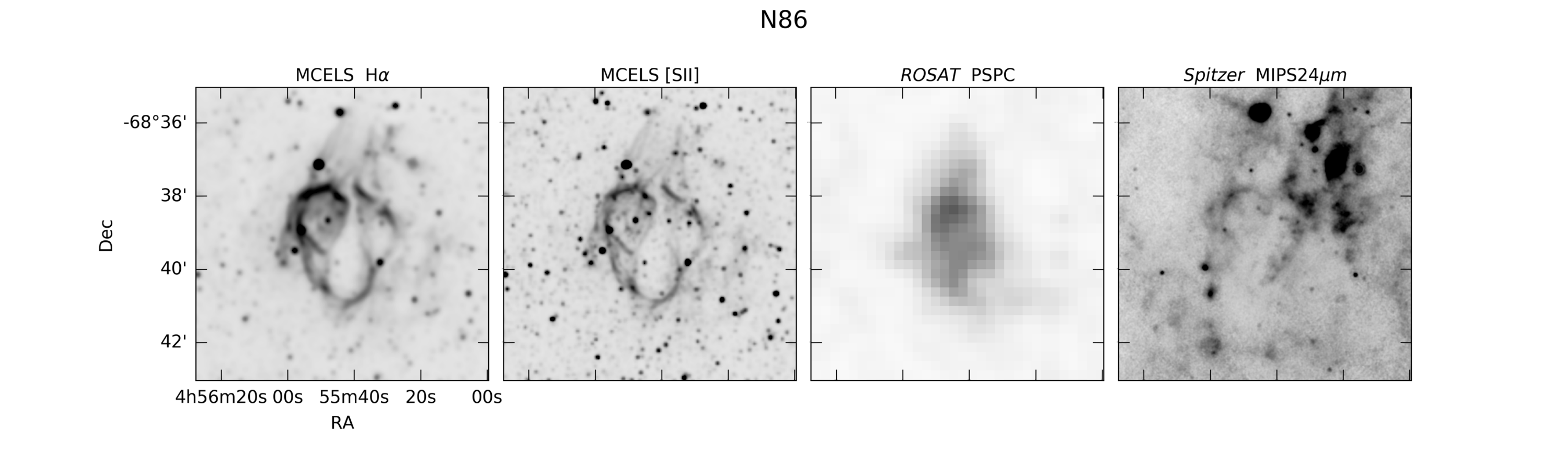}
\includegraphics[scale=0.25]{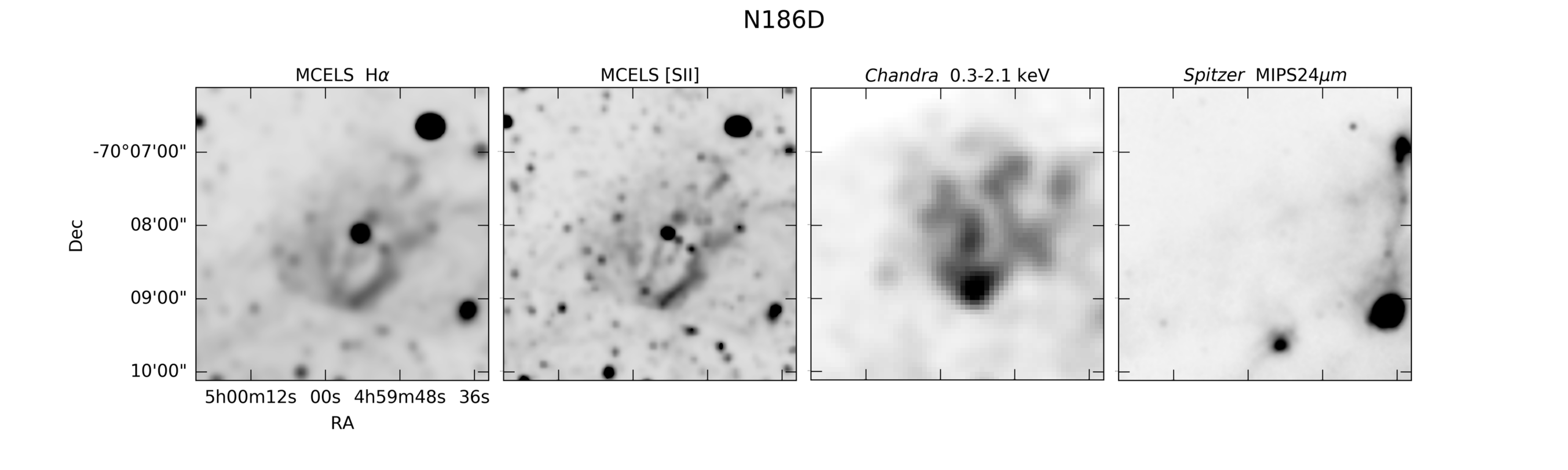}
\includegraphics[scale=0.25]{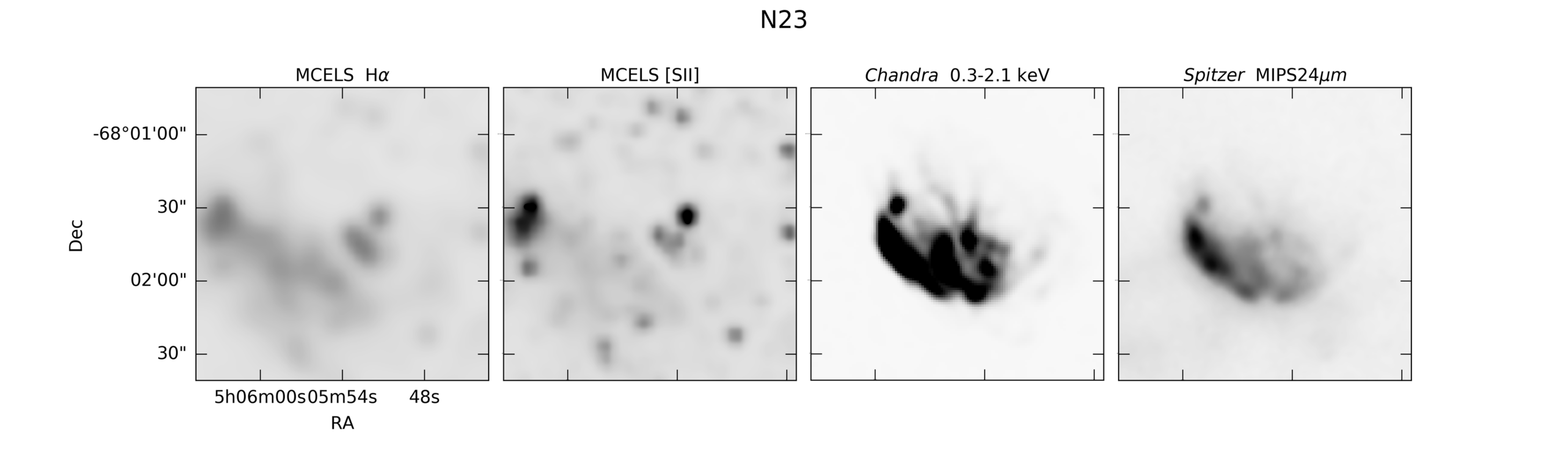}
\includegraphics[scale=0.25]{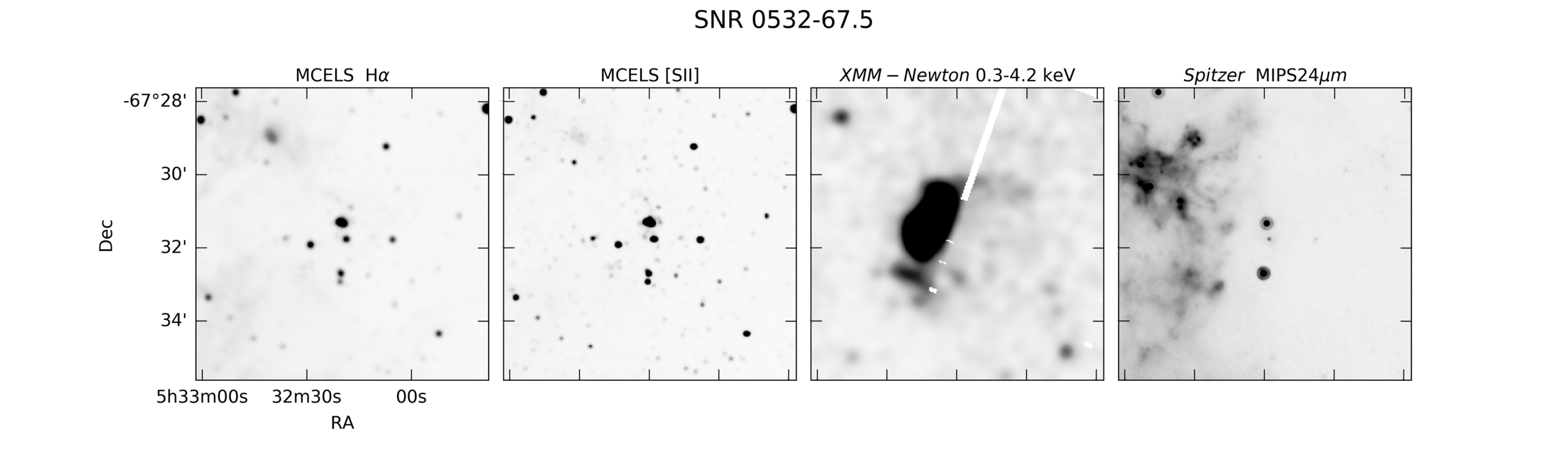}
\includegraphics[scale=0.25]{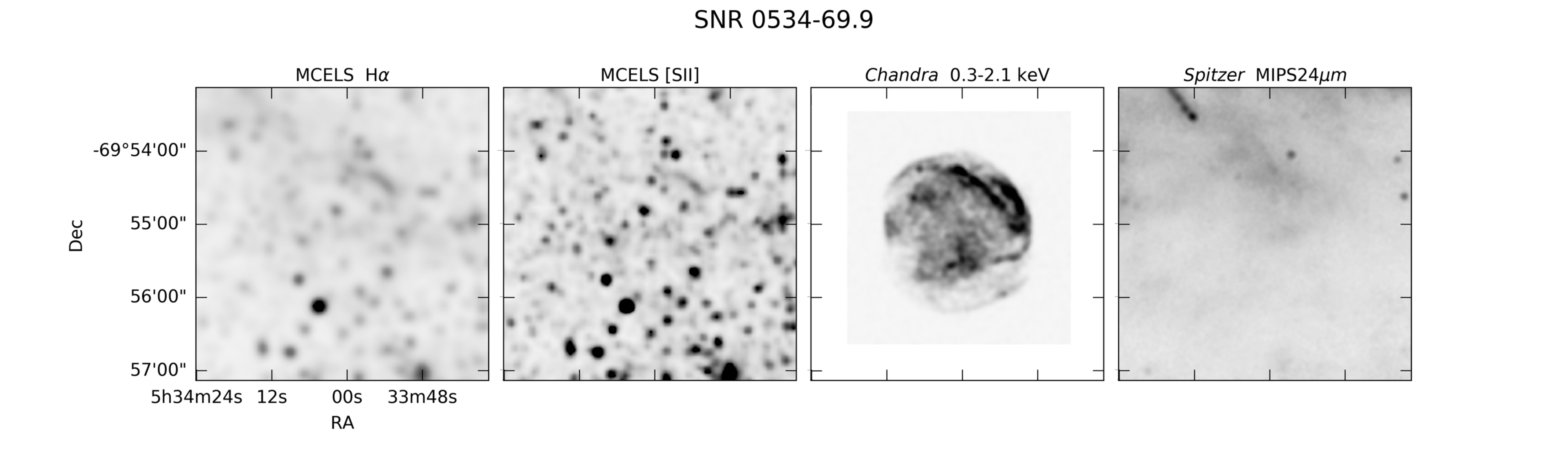}
\end{figure*}

\begin{figure*}[h]
\centering
\includegraphics[scale=0.25]{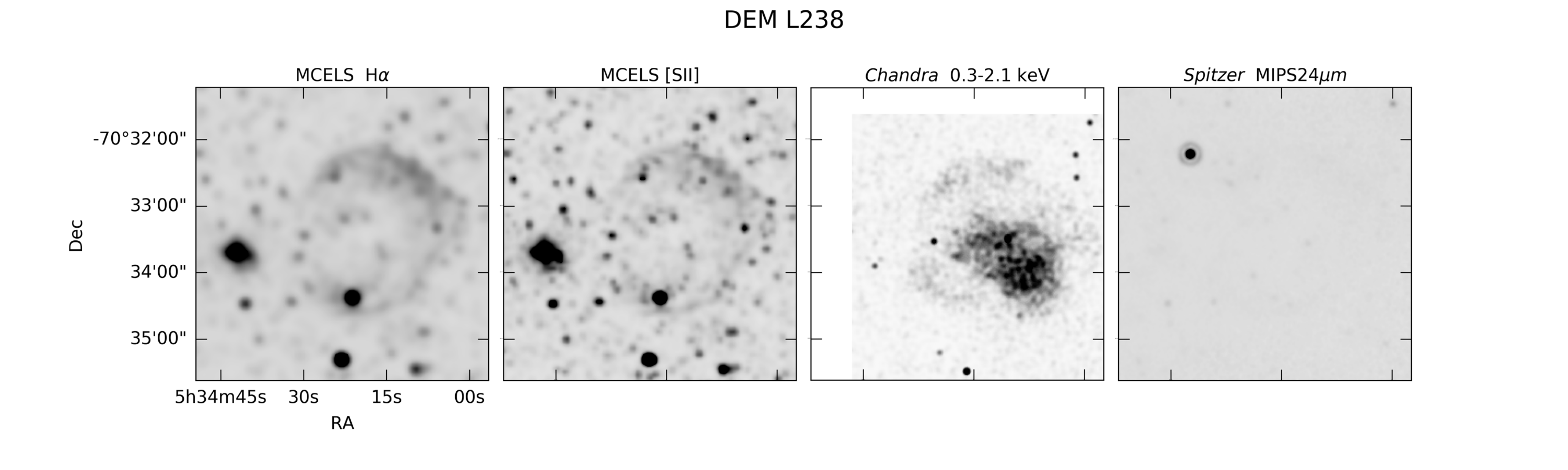}
\includegraphics[scale=0.25]{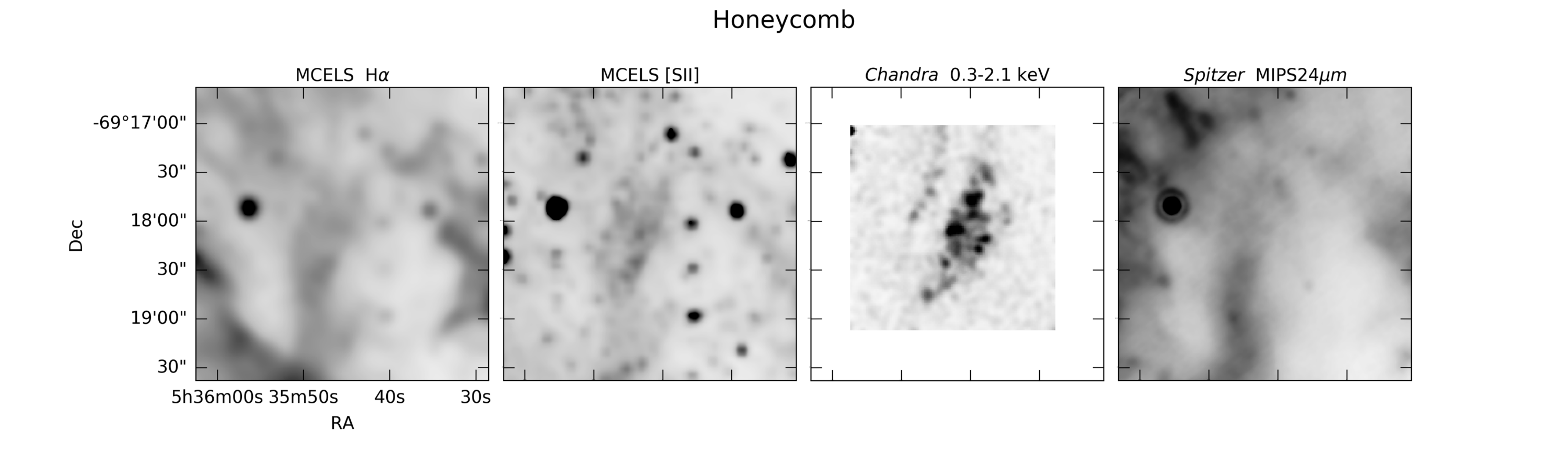}
\includegraphics[scale=0.25]{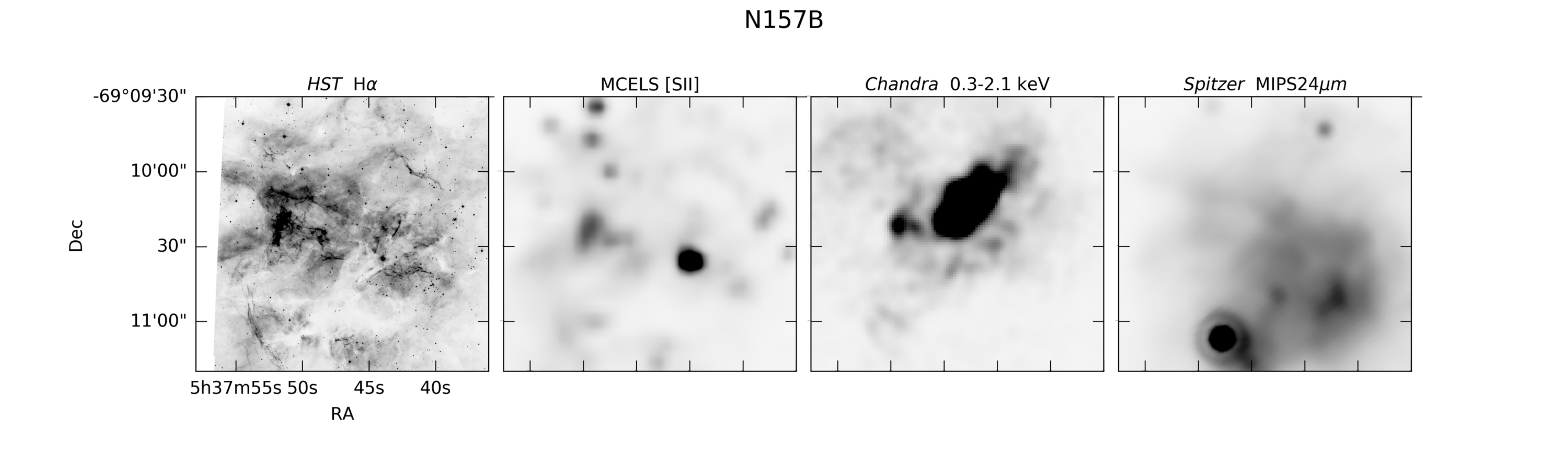}
\includegraphics[scale=0.25]{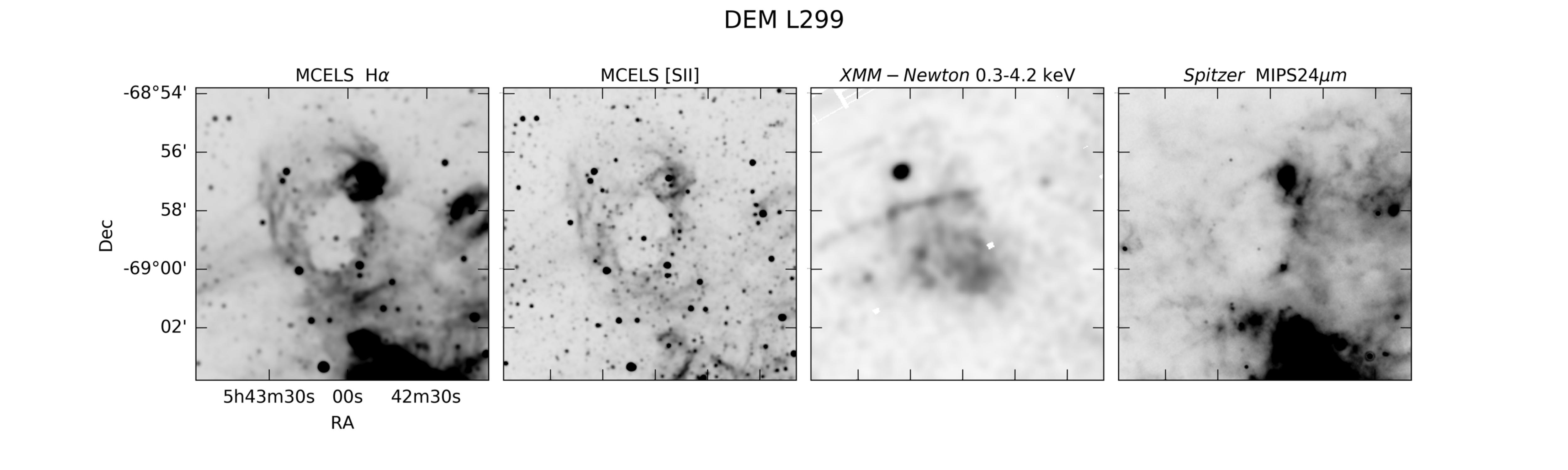}
\includegraphics[scale=0.25]{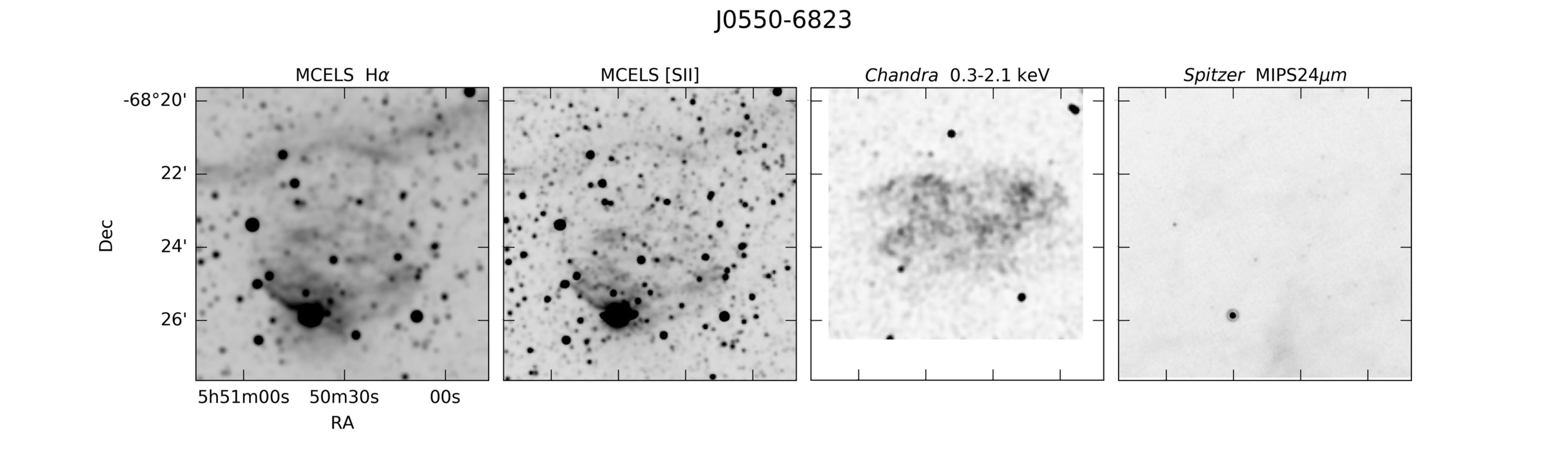}
\caption{Images of the SNRs with large discrepancies between the size measurements of Bo2017 and De2010.
}
\end{figure*}


\clearpage
\section{Descriptions of size determinations for individual SNRs}
\emph{N86}.-- The shape of the SNR is irregular because of the breakout structure in the north, through which the hot gas flows out from the SNR shell \citep{williams1999}.  For N86 in Figure 2, we have used both sizes of De2010 (75.0 pc, including the breakout) and Bo2017 (61.5 pc, not including the breakout) in the $L_X$ -- size plot to illustrate the uncertainty in the size. Only low-resolution \emph{ROSAT} X-ray images are available for N86; high-resolution \emph{Chandra} and \emph{XMM-Newton} observations will help refine the size determination.

\emph{N186D}.-- The size of SNR N186D in H$\alpha$ cannot be unambiguously measured, because N186D is projected on the rim of the superbubble N186E.  Through the analysis of velocity fields in N186D, \citet{laval1989} determined the SNR size to be $\sim$40 pc, which is consistent with the size of the [\ion{S}{2}]-enhanced shell (see Figure 5 in Appendix A).  Based on these considerations, we adopt the size 36.8 pc given by De2010. 

\emph{N23}.-- X-ray emission of N23 is enhanced in the southeast side, likely due to a denser ambient medium \citep{williams1999}. The size reported by De2010, $18 \times 12$ pc, corresponds to only the X-ray-brightest region.  \citet{maggi2016} and Bo2017 included the fainter X-ray emission from the northwest side and reported a larger SNR size, 23.6 pc, which is more accurate and hence adopted in the $L_X$--size plot in Figure 2. 

\emph{SNR 0532-67.5}.-- This SNR may be associated with the OB association LH75 \citep{chu1997}.  This SNR has no optical counterpart, indicating that it is in a low-density medium, possibly caused by the fast stellar winds and SN explosions from LH75.  The size of this SNR can be measured only in X-rays. There is bright X-ray emission in a $\sim 40 \times 20$ pc region, and a fainter and larger X-ray arc connected with the bright region. As in the case of the SNR N23, we include both the bright and faint X-ray emission regions in the size estimate, and adopt a size of 67.5 pc as the size of SNR 0532-67.5.

\emph{SNR 0534-69.9}.-- The optical images of this SNR show only a faint filament associated with the brightest X-ray emission region.  \emph{Chandra} observations show the SNR clearly in X-rays, although the southern rim is much fainter than the rest of the SNR.  We have measured and adopted the full extent of the SNR shown in X-rays, about 33.5 pc, similar as the size measured by \citet{maggi2016},
which is larger than those reported by De2010 and Bo2017.

\emph{DEM L238}.-- Comparing H$\alpha$ and \emph{Chandra} images, the X-ray emitting region is larger than the optical shell. We adopt the full extent of the SNR, 47.5 pc.

\emph{Honeycomb}.-- The Honeycomb SNR is near the 30 Doradus complex, and to the south of the superbubble 30 Dor C.  This region has a very complex star formation history and chaotic nebular morphology.  The lack of bright ionized gas region suggests an evolved environment with low ISM densities.  The optical morphology of the SNR is very irregular, consisting of many cells instead of a simple shell \citep{chu1995, meaburn2010}, leading to large uncertainty in the determination of SNR size.  For the Honeycomb SNR, we have used both sizes of De2010 (15 pc) and Ba2010 (25.5 pc) in the $L_X$ -- size plot to illustrate the uncertainty in the size.

\emph{N157B}.-- The environment of N157B in H$\alpha$ is very complex because this SNR is superposed on the HII region of the OB association LH99 \citep{chu1997}, and dissected by a foreground dark cloud.  The most reliable measurement of the SNR size is through the analysis of gas kinematics using long-slit high-dispersion spectroscopic observations, 25$\times$18 pc \citep{chu1992}. 
This SNR boundary has been confirmed by sharp filaments revealed by \emph{HST} images as shown in Figure 5.
The size 21.8 pc given by De2010 is taken from \citet{chu1992}. 

\emph{DEM\,L299}.-- This SNR is inside a large optical shell.  The size reported by De2010 corresponds to the large optical shell.  The X-ray emission actually extends from the shell cavity to the southwest, indicating an outflow.  The [\ion{S}{2}]/H$\alpha$ ratio is enhanced in the shell structure and in the superposed filaments of supergiant shell LMC-2.  The SNR is clearly in a very complex environment.  We include all the diffuse X-ray emission region and [\ion{S}{2}] enhanced filaments, and measure a size of 100$\times$50 pc.  A smaller SNR size, $\sim$55 pc,  has been reported by \citet{warth2014} and \citet{maggi2016} based on the diffuse X-ray emission and a surrounding [\ion{S}{2}]-enhanced filament.  The large discrepancy between these two size measurements illustrate the difficulty in determining SNR sizes in a complex environment  confused by  other energetic feedback processes from massive stars. We adopt both 73.5 and 56.5 pc in Table 1 (Appendix C) and Figure 2.

\emph{J0550-6823}.-- While the diameter of the optical shell is only $\sim$68 pc, there is X-ray and radio emission extending over 90 pc in the east-west direction; therefore, we adopt the size of 90$\times 68$ pc by \citet{bozzetto2012}.

\clearpage
\section{Sizes and X-ray luminosities of LMC SNRs}
\startlongtable
\begin{deluxetable}{ccccccccc}
\tablecaption{Sizes and X-ray luminosities of LMC SNRs}
\tablehead{\colhead{SNR J2000 \tablenotemark{a}}&\colhead{Other Name} & \colhead{Size (Ba2010)} & \colhead{Size (De2010)}& \colhead{Size (Bo2017)}& \colhead{Large Discrepancy \tablenotemark{b}}&\colhead{Adopted Size} & \colhead{$L_{\textrm x}$\tablenotemark{c}} \\ 
\colhead{} & \colhead{} & \colhead{(pc)} & \colhead{(pc)} & \colhead{(pc)} & \colhead{($>$16$\%$)} & \colhead{(pc)} & \colhead{(10$^{35}$ergs/s)} } 
\startdata
J0448-6700 & [HP99] 460 & 55.0 & 59.2 &60.8&& 60.8 & 0.46 \\
J0449-6920 & --& 33.2 &30.0& 28.8 &  & 28.8 & 0.07 \\
J0450-7050 & SNR 0450-70.9& 89.2 & 97.5 & 109.4 &  & 109.4 & 0.59 \\
J0453-6655 & SNR in N4 & 63.0& 64.5 & 60.6&  & 60.6 & 1.17 \\
J0453-6829 & SNR 0453-68.5 & 30.0& 30.0& 30.4&  & 30.4 & 13.85 \\
J0454-6713 & SNR 0454-67.2 & 44.2 & 37.5& 32.5 &  & 32.5 & 1.58 \\
J0454-6626 & N11L & 21.8 & 18.0& 20.4 &  & 20.4 & 0.63 \\
J0455-6839 & N86 & 87.0  & 75.0 & 61.5 & Y & 61.5--75.0 & 1.42\\
J0459-7008 & N186D & 37.5 & 36.8 & 29.0 & Y & 36.8 & 1.09 \\
J0505-6753 & DEM L71 &18.0  & 20.2 & 18.6 && 18.6 & 44.59 \\
J0505-6802 & N23 & 27.8  & 15.0 & 23.6& Y & 23.6 & 26.25 \\
J0506-6541 & DEM L72 & 102.0 &83.2 & 96.2&& 96.2 & 0.53  \\
J0506-7026 & [HP99] 1139 & 82.5 & -- & 42.5 && 42.5 & 1.44  \\
J0508-6902 & [HP99] 791 & -- & -- & 67.0&& 67.0 & 0.37 \\
J0508-6830 & --& -- & -- & 30.8&  & 30.8 &0.09 \\
J0509-6844 & N103B & 7.0 & 7.5 & 7.0 && 7.0 & 51.7 \\
J0509-6731 & SNR 0509-67.5 & 7.25 &8.4 & 7.6 && 7.6 & 16.51 \\
J0511-6759 & --& -- & -- & 55.5& & 55.5 & 0.16 \\
J0512-6707 & [HP99] 483 & -- & -- & 12.5&& 12.5 & 0.09 \\
J0513-6912 & DEM L109 & 53.8 & 57.8 & 55.6 && 55.6 & 0.51 \\
J0514-6840 & -- & -- & -- &55.0&& 55.0 & 0.4 \\
J0517-6759 & -- & -- & -- & 66.8 && 66.8 & 0.24 \\
J0518-6939 & N120 & 33.5 & 21.8 & 23.4 && 23.4 & 0.88 \\
J0519-6902 & SNR 0519-69.0 & 7.8 & 8.2 & 8.6 && 8.6 & 34.94 \\
J0519-6926 & SNR 0520-69.4 & 43.5 & 33.8 & 31.2 && 31.2 & 2.69 \\
J0521-6543 & DEM L142 & -- & 40.5 & 34.5 && 34.5 & -- \\
J0523-6753 & SNR in N44 & 57.0 & 52.5 & 57.5 && 57.5 &0.9 \\
J0524-6624 & DEM L175a & 58.5 & 51.8 & 36.25 && 36.2&-- \\
J0525-6938 & N132D & 28.5 &  26.2& 25.5 && 25.5 & 315.04 \\
J0525-6559 & N49B & 42.0 & 36.0 & 38.8 && 38.8 & 38.03 \\
J0526-6605 & N49 & 21.0 & 21.0 & 18.8 && 18.8 & 64.37 \\
J0527-6912 & SNR 0528-69.2 & 36.8 & 35.2 & 35.0 && 35.0 & 1.99 \\
J0527-6550 & DEM L204 & 75.8 & 67.5& 76.2 && 76.2 & -- \\
J0527-6714 & SNR 0528-6716 & -- & -- & 54.0 && 54.0 & 0.58  \\
J0527-7104 & [HP99] 1234 & 49.0 & -- & 70.2 && 70.2 & 0.25  \\
J0528-6727 & DEM L205 & -- & -- & 55.0 & & 55.0 & 0.21  \\
J0529-6653 & DEM L214 & 25.0 & -- & 33.1 && 33.1 & 1.04 \\
J0530-7008 & DEM L218 & 53.2 & 47.2 & 49.4 && 49.4 &0.72 \\
J0531-7100 & N206 & 48.0 & 45.0& 45.0 &  & 45.0 & 2.55  \\
J0532-6732 & SNR 0532-67.5 & 63.0 & 67.5 & 45.0 & Y & 67.5 & 2.48 \\
J0533-7202 & -- & -- & -- & 45.0 && 45.0 & 0.57 \\
J0534-6955 & SNR 0534-69.9 & 28.5 & 23.2 & 28.8 & Y & 33.5 & 6.33 \\
J0534-7033 & DEM L238 & 45.0 & 40.5 & 47.5 & Y & 47.5 & 1.55 \\
J0535-6916 & SN 1987A & 0.5 & $>$1.5 & 0.45 && 0.45 & 27.39 \\
J0535-6602 & N63A & 16.5 & 19.5 & 18.5 && 18.5 & 185.68 \\
J0535-6918 & Honeycomb & 25.5 & 15.0 & 18.8 & Y & 15.0--25.5 & 0.4  \\
J0536-6735 & DEM L241 & 33.8 & 36.0 & 34.0 & & 34.0 & 3.84  \\
J0536-7039 & DEM L249 & 45.0 & 37.5 & 39.2 && 39.2 & 1.43 \\
J0536-6913 & SNR 0536-69.2 & 120 & -- & 16.5 & & 16.5 & 0.22 \\
J0537-6628 & DEM L256 & 51.0  & 48.0 & 46.9 && 46.9 & 0.32 \\
J0537-6910 & N157B & 25.5 & 21.8 & 31.5 & Y & 21.8 & 15.0 \\
J0540-6944 & SNR in N159 & 19.5 & 27.0& 26.2 && 26.2 & 0.43 \\
J0540-6920 & SNR 0540-69.3 & 15.0 & 18.0& 15.6 && 15.6 & 87.35 \\
J0541-6659 & [HP99] 456& -- & -- & 71.5 && 71.5 & 0.77 \\
J0543-6858 & DEM L299 & 79.5 & 73.5 & 56.5 & Y & 56.5--73.5 & 1.68 \\
J0547-6943 & DEM L316B & 21.0 & 46.5 & 45.0 && 45.0 & 1.47 \\
J0547-6941 & DEM L316A & 14.0 & 30.0& 30.0 && 30.0 & 1.26 \\
J0547-7025 & SNR 0548-70.4& 25.5 & 28.5 & 28.1 && 28.1 & 2.94 \\
J0550-6823 & -- & 78.0 & 65.2& 81.9 & Y & 81.9 & 1.22 \\
\enddata
\tablenotetext{a}{The 59 confirmed SNRs listed in \citet{maggi2016}.}
\tablenotetext{b}{Size(De2010)/Size(Bo2017)$>$1.16 or $<$0.86.}
\tablenotetext{c}{Taken from \citet{maggi2016}.}
\end{deluxetable}

\end{CJK*}
\hspace{1cm}
\end{document}